\documentclass[a4paper,11pt]{article}
\usepackage{amsmath}
\usepackage{amsfonts}
\usepackage{amssymb}
\usepackage{bbm}
\usepackage{geometry}
\usepackage[parfill]{parskip} 
\usepackage{graphicx}
\usepackage{epstopdf}
\usepackage[colorlinks=true, pdfstartview=FitV, linkcolor=blue, citecolor=blue, urlcolor=blue]{hyperref}

\begin{document}

\def\FF{\mathbbm{F}}
\def\RR{\mathbbm{R}}
\def\CC{\mathbbm{C}}
\def\QQ{\mathbbm{Q}}
\def\ZZ{\mathbbm{Z}}
\def\NN{\mathbbm{N}}
\def\Pr{\mathbbm{P}}
\def\Ex{\mathbbm{E}}

\def\H{\mathbf{H}}

\def\IQP{\mathbf{IQP}}
\def\BPP{\mathbf{BPP}}
\def\PP{\mathbf{PP}}
\def\P{\mathbf{P}}
\def\L{\mathbf{L}}
\def\BQP{\mathbf{BQP}}

\def\cC{\mathcal{C}}
\def\cL{\mathcal{L}}
\def\cM{\mathcal{M}}
\def\cP{\mathcal{P}}
\def\cQ{\mathcal{Q}}

\def\ket#1{|\,#1\,\rangle}
\def\bra#1{\langle\, #1\,|}
\def\braket#1#2{\langle\, #1\,|\,#2\,\rangle}
\def\ketbra#1#2{\ket{#1}\bra{#2}}
\def\identity{\leavevmode\hbox{\small1\kern-3.8pt\normalsize1}}
\def\span#1{\left< #1 \right>}

\def\X{\mathbf{X}}
\def\a{\mathbf{a}}
\def\b{\mathbf{b}}
\def\c{\mathbf{c}}
\def\m{\mathbf{m}}
\def\k{\mathbf{k}}
\def\x{\mathbf{x}}
\def\y{\mathbf{y}}
\def\p{\mathbf{p}}
\def\s{\mathbf{s}}
\def\t{\mathbf{t}}
\def\d{\mathbf{d}}
\def\e{\mathbf{e}}
\def\0{\mathbf{0}}
\def\1{\mathbf{1}}

\newtheorem{theorem}{Theorem}
\newtheorem{corollary}{Corollary}[theorem]
\newtheorem{conjecture}{Conjecture}
\newtheorem{definition}{Definition}
\newtheorem{lemma}{Lemma}
\newtheorem{propos}{Proposition}
\newtheorem{example}{Example}
\newcommand{\qed}{\mbox{\rule[0pt]{1.5ex}{1.5ex}}}

\title{Binary Matroids and Quantum Probability Distributions}

\author{Dan Shepherd\footnote{djs\_at\_cantab.net}}

\maketitle

\begin{abstract}
  We characterise the probability distributions that arise from quantum circuits all of whose gates commute, and show when these distributions can be classically simulated efficiently.  We consider also marginal distributions and the computation of correlation coefficients, and draw connections between the simulation of stabiliser circuits and the combinatorics of representable matroids, as developed in the 1990s.
\end{abstract}

\subsubsection*{Acknowledgements}

Much of this research was undertaken at the University of Bristol, in context of related work with Richard Jozsa and Michael Bremner (\emph{cf}. \cite{us:BJS}).  The research was primarily funded by CESG.

\section{Introduction}  \label{sect:intro}

  It is widely held that quantum computation is a useful paradigm for algorithmics because it enables us to describe a wider class of algorithms than can be described classically, and some of these may represent qualitatively faster solutions to certain problems than could ever be obtained classically, in principle.  To put this sentiment on a more rigourous footing, one often sees the conjecture phrased in terms of classes of decision languages circumscribed by asymptotic time constraints, \emph{e.g.} \emph{Conjecture :} $\BQP \not= \BPP$.  A more general approach looks not at \emph{decision languages} arising from uniform families of circuits, but at \emph{uniform families of probability distributions} arising from uniform families of circuits.  Then instead of asking ``Can this decision language be computed efficiently?'' one asks more general questions such as ``Can this probability distribution be sampled from efficiently?  Can this probability distribution be \emph{approximately} sampled from efficiently?  Can we efficiently compute the correlation coefficients of this distribution?  Given samples from the distribution, can we efficiently verify the hypothesis that they were thus sampled?'' and so on.  The reader is directed to \cite{us:BJS} for more complexity-theoretic background and analysis on questions of this type.
  
  In this paper, we restrict attention to the combinatorial analysis of a particular kind of uniform family of quantum circuits called $\IQP$, first introduced in \cite{me:IQC} and discussed at greater length in \cite{mythesis}.  We call such circuits \emph{X-programs} (see \S\ref{sect:main}) because they are described by Hamiltonians composed entirely of Pauli $X$ operators.  From the perspective of implementation, these circuits can be rendered by any model for computing in the class $\mathbf{QNC}_f^0$ (defined in \cite{lit:Hoy02}).  In this paper (\S\ref{sect:main}), we fully describe the probability distributions that they generate, and also give formulas for the correlation coefficients associated to these distributions.  We show when these probabilities and correlations are computable efficiently classically, and when they are not.  We recall the argument \cite{mythesis,us:BJS} for why it is generally impossible to sample efficiently classically from these distributions, unless the polynomial hierarchy collapses.  This is  by no means the first time a fully combinatoric treatment has been made of physical problems (\emph{e.g.} \cite{lit:Wel90}).
  
  All of these results depend on describing X-programs in terms of binary matroids; equivalently binary linear codes.  We begin by recalling the relationship between matroids and codes, and we note a link between the efficient simulation of Clifford circuits \cite{lit:AG0406} and Vertigan's algorithm for binary matroids \cite{lit:Vert98}.  Our aim is to provide a fairly self-contained reference, so all of the combinatorial definitions needed will be given up front in \S\ref{sect:Comb}, where (almost) no mention will be made of quantum computation.

\section{Combinatorics}  \label{sect:Comb}

  This section introduces the necessary combinatorics notation and ideas.  We give three propositions, together with sketch proofs.  To proceed to the results of \S\ref{sect:main}, the uninitiated reader will want to understand the statements of these propositions, though perhaps not necessarily learn the methods of their proofs.  The key subsection here is \S\ref{sect:complexity}, where the main computational complexity results are recalled.  In \S\ref{sect:trans} we give worked examples of the transformations that are used later.
  
  Let $P$ be any binary matrix, that is, any matrix over $\FF_2$ having, say, $n$ rows and $l$ columns and rank $r$.  To it we associate a \emph{binary linear code} $\cC = \cC(P)$ and a \emph{binary matroid} $\cM = \cM(P)$.  The code $\cC$ will be generated by the columns of $P$, and the matroid $\cM$ will be coordinatized by the rows of $P$.  Such codes and matroids are in one-to-one correspondence (Proposition~\ref{propos:121} below), so everything that can be said about the one can be said about the other~: accordingly we will try to say everything twice, for clarity.

\subsection{Formal definitions}  \label{sect:defs}
  
  Formally, $\cC$ is defined to be the subset of the vector space $\FF_2^n$ that is generated by linearly combining \emph{columns} of $P$.  The cardinality of the code is thus $2^r$, so we call $r$ its \emph{rank}.  The elements of $\cC$ are called \emph{codewords}, and these may also be thought of as binary strings of length $n$.  The parameter $n$ is called the \emph{length} of the code.  The parameter $l$ is not well-defined for the code, only for its generator matrix $P$, though clearly $r \le l$.
  
  Formally, $\cM$ is defined to be the isomorphism class of the (multi)set $E$ of rows of $P$ (vectors in the vector space $\FF_2^l$) together with the induced rank function $\rho_\cM$ that maps subsets of these row-vectors to their rank, \emph{i.e.} to whichever integer counts the dimension of the subspace of $\FF_2^l$ that they span.  In particular, $\rho_\cM(\emptyset) = 0$ and $\rho_\cM(E) = r$, by definition.  The parameter $n$ is called the \emph{size} of the matroid, and $r$ is called its \emph{rank}.  The parameter $l$ is not well-defined for the isomorphism class, since the row-vectors could easily be embedded in a larger vector space, though clearly $r \le l$.
  
\begin{propos}  \label{propos:121}
  For binary matrices $P$ and $Q$, the following are equivalent~:
\begin{itemize}
  \item  $P$ and $Q$ generate the same code, $\cC(P) = \cC(Q)$;
  \item  $P$ and $Q$ generate the same matroid, $\cM(P) = \cM(Q)$;
  \item  there exist binary matrices $R$ and $R'$ such that $P = Q \cdot R$ and $P \cdot R' = Q$.
\end{itemize}
\end{propos} 

  To prove this, we give a \emph{normal form} for a matrix $P$.  First, identify a basis for the matroid $\cM(P)$, that is, find $r$ rows that collectively have rank $r$.  Without loss of generality, \emph{i.e.} for illustration, we assume these to be the first $r$ rows of $P$.  Then we can write
\begin{eqnarray*}
P  &=&  \left( \begin{array}{c} B \\ C \end{array} \right)
  ~~=~~ \left( \begin{array}{c} I \\ D \end{array} \right) \cdot B, \\
P' &:=& \left( \begin{array}{c} I \\ D \end{array} \right),
\end{eqnarray*}
where $B$ is the submatrix given by the first $r$ rows, and $C$ is the remaining submatrix.  Here $I$ is an $r$-by-$r$ identity matrix.  We can find $D$ uniquely satisfying $D\cdot B = C$ precisely because $B$ is a basis for the space to which every row of $P$ belongs.  This kind of reduction can be achieved by \emph{Gaussian elimination}, and $P'$ is called an \emph{echelon reduction} of $P$.  Echelon reduction is unique once an ordered basis is chosen, but we are free to choose any basis with any ordering.  Elsewhere in the paper we use the notation $P'$ to denote an echelon reduction of $P$.  

The reader may complete the proof of the Proposition, using this idea of identifying a basis set and an echelon reduction.  
\hfill  \qed

\subsection{Weight enumerator and Tutte polynomial}  \label{sect:WETP}

  The weight enumerator polynomial of $\cC$ is a monovariate polynomial (\emph{i.e.} element of $\ZZ[\zeta]$ for indeterminate $\zeta$) given by
\begin{eqnarray*}
  W_{\cC}( \zeta )  &:=&  \sum_{\c \in \cC} \zeta^{|\c|},
\end{eqnarray*}
where $|\c|$ denotes the Hamming weight of the codeword $\c$.

  The Tutte polynomial of $\cM$ is a bivariate polynomial (\emph{i.e.} element of $\ZZ[x,y]$ for indeterminates $x,y$) given by 
\begin{eqnarray*}
  T_{\cM}( x, y )  &:=&  
    \sum_{ X \subseteq E } (x-1)^{\rho_\cM(E) - \rho_\cM(X)} \cdot 
                           (y-1)^{|X|-\rho_\cM(X)},
\end{eqnarray*}
where $\rho_\cM : E \rightarrow \ZZ$ is the rank function defining the matroid.

Here is a random example, with $n=6$, $r=3$, $l=4$~:
\begin{eqnarray*}
  P ~=~ \left( \begin{array}{cccc} 
    1&1&0&1\\0&1&1&0\\0&0&0&0\\0&1&0&1\\1&0&1&1\\0&1&0&1
  \end{array} \right),  
  &&
  P' ~=~ \left( \begin{array}{ccc} 
    1&0&0\\0&1&0\\0&0&0\\0&0&1\\1&1&0\\0&0&1
  \end{array} \right); \\
  W_{\cC(P)}(\zeta)  &=&  1 + 4\zeta^2 + 3\zeta^4; \\
  T_{\cM(P)}(x,y)    &=&  y(x+y)(x^2+x+y).
\end{eqnarray*}

The natural connection between these two concepts is given by Greene's theorem~:
\begin{propos}{(C. Greene, 1976)}  \label{propos:Greene}
\begin{eqnarray*}
  W_{\cC}( \zeta )  &=&  
    \zeta^{n-r} \cdot (1-\zeta)^{r} \cdot 
    T_{\cM}\left( \frac{1+\zeta}{1-\zeta}, \frac1\zeta \right),
\end{eqnarray*}
where $n = |E|$ and $r=\rho_\cM(E)$ and $E$ is the base set for $\cM$, as usual.
\end{propos}

  This can be proved inductively, using the entirely standard matroid notions of \emph{deletion} and \emph{contraction}, and taking for the base cases those few examples where $r \le n \le 1$.  Note that \emph{loops} and \emph{coloops} are conveniently dealt with first as special cases.
\hfill  \qed

  Throughout this paper, we find it convenient to use a \emph{normalised} version of the weight enumerator/Tutte polynomial, given in terms of a single \emph{real} parameter $\theta$.  We fix the following notation for use throughout the rest of the paper :
\begin{eqnarray*}
  \alpha_{(P,\theta)}  
    &:=&  2^{-r} \cdot e^{i \theta n} \cdot 
          W_{\cC(P)}\left(~ e^{-2i \theta} ~\right) \\
    &=&   e^{i \theta (r-n)} \cdot i^r \cdot \sin^r \theta \cdot
          T_{\cM(P)}\left(~ \frac{e^{i\theta}+e^{-i\theta}}{e^{i\theta}-e^{-i\theta}}, ~e^{2i\theta} ~\right),
\end{eqnarray*}
again with $r$ for the rank and $n$ for the length (size) of the code (matroid).  Note that $\theta$ is really only defined modulo $2\pi$, and inverting the sign of $\theta$ only conjugates the output $\alpha$ value trivially.

Here is a useful observation, showing how normalisation ensures $|\alpha_{(P,\theta)}| \le 1$~:
\begin{propos}  \label{propos:alpha}
  The scalar $\alpha_{(P,\theta)}$ defined above can be expressed directly in terms of the code $\cC(P)$ according to the probabilistic formula below.
\begin{eqnarray*}
  \alpha_{(P,\theta)} 
    &=&  \Ex_{\c \in \cC(P)}
         \left[~ \exp\bigl( i\theta \cdot ( n - 2|\c| ) \bigr) ~\right].
\end{eqnarray*}
\end{propos}
To see this, observe that the factor $e^{i \theta n}$ can be pulled out in front of the expectation operator, and what is left constitutes a probabilistic interpretation of the required value $2^{-r} \cdot W_\cC( e^{-2i\theta})$, because there are $2^r$ codewords in the code.
\hfill  \qed

\subsection{Computational Complexity}  \label{sect:complexity}

What is the computational complexity of computing the value of the scalar $\alpha_{(P, \theta)}$?  This question is addressed in \cite{lit:JVW90} and \cite{lit:Vert98}.  The multiples of $\frac\pi4$ are found all to be essentially trivial.  

For example, if $\theta = \pi$ then $\alpha_{(P, \theta)} = (-1)^n$.  
If $\theta = \frac\pi2$ then $\alpha_{(P, \theta)}$ vanishes whenever $\cC$ is not an even code, and evaluates to $i^n$ whenever it is even.  One can determine whether a code is even by looking at \emph{any} generator matrix for it~: the code is not even iff any column of $P$ has odd Hamming weight.
If $\theta = \frac\pi4$ then Vertigan's algorithm \cite{lit:Vert98} provides an efficient explicit polynomial-time recursion to evaluate $T_\cM(-i,i)$, which is proportional to $\alpha_{(P, \frac\pi4)}$.
This algorithm is remarkably similar to the algorithm used in classically simulating the probability distribution associated to a Gottesman-Knill-Clifford computation \cite{lit:AG0406}.

Conversely, for any other values of $\theta$, the worst-case complexity for computing $\alpha_{(P, \theta)}$ (over the class of all binary matrices $P$) is $\mathbf{\#P}$-hard, and there are efficient reductions from one $\theta$ to any other (excluding multiples of $\frac\pi4$, limiting to algebraic values of $e^{i\theta}$) \cite{lit:JVW90,lit:Vert98}.  In particular, the reader should note that $\theta = \frac\pi8$ is no less hard than any other value of $\theta$, when it comes to evaluating $\alpha_{(P, \theta)}$ in the worst case.  

The same hardness holds for specific subclasses of matroid, \emph{e.g.} for \emph{graphic} matroids~\cite{lit:JVW90}.  But it should come as no surprise that there are `trivial' classes of matroid where $\alpha_{(P, \theta)}$ is readily evaluated for all $\theta$.  The primary example would seem to be the case of graphic matroids having \emph{bounded treewidth} \cite{lit:And95}.
Jaeger \emph{et al.} also claim that $T_\cM(x,y)$ is easily evaluated when $(x-1)(y-1)=2$ (as is always the case for our $\alpha_{(P,\theta)}$) whenever $\cM$ is the cycle matroid of a \emph{planar} graph (\emph{cf.} \cite{lit:JVW90} \S(5.8), also \cite{lit:Vert06} lemma~12.2, and for the main reduction, see \cite{lit:Fish66}).

\subsection{Transformations}  \label{sect:trans}

There are two matroid/code transformations that we shall be requiring in \S\ref{sect:main} for our first two theorems.  Notation for these is given here, but is not generally found in the literature.  We include example computations of both transformations, to add clarity.

\subsubsection*{Projection}
\def\proj#1#2{#1\top#2}

Let $\x$ be a non-zero string of $l$ bits, and from the matrix $P$ (\emph{resp.} the code $\cC$, the matroid $\cM$) we derive $\proj{P}\x$ (\emph{resp.} $\proj\cC\x$, $\proj\cM\x$) by projecting each row along direction $\x$ onto a hyperplane avoiding $\x$.  This projection operation is well-defined for both $\cC$ and $\cM$, regardless of which hyperplane is chosen.  

To make it well-defined for $P$ as well, we specify (somewhat arbitrarily) that this projection is achieved by mapping a row $\a$ to whichever of $\{\a,\a+\x\}$ is lexicographically first.
Note that the rank of $\proj\cC\x$ (and of $\proj\cM\x$) will either be one less than that of $\cC$ (and $\cM$), or else $\proj\cC\x = \cC$ (and $\proj\cM\x = \cM$).  In either case, $\proj\cC\x \subseteq \cC$ and the length is not changed.
Note that if $\x$ was originally a row of $P$, then $\proj{P}\x$ contains an all-zero row, so the matroid $\proj\cM\x$ will contain a loop.

Here is a random example, with $n=6$, $r=3$, $l=4$, and $\x = (0110)$~:
\begin{eqnarray*}
  P ~=~ \left( \begin{array}{cccc} 
    1&1&0&1\\0&1&1&0\\0&0&0&0\\0&1&0&1\\1&0&1&1\\0&1&0&1
  \end{array} \right)
  &\rightarrow&
  \proj{P}\x ~=~ \left( \begin{array}{cccc} 
    1&0&1&1\\0&0&0&0\\0&0&0&0\\0&0&1&1\\1&0&1&1\\0&0&1&1
  \end{array} \right)
\end{eqnarray*}
In this example, the code/matroid is indeed changed and so the rank drops by one from 3 to 2.  One more loop is created (second row) and one more non-trivial parallel class is created (first and fifth rows now parallel).
Echelon-reduced forms of these example matrices are
\begin{eqnarray*}
  P' ~=~ \left( \begin{array}{ccc} 
    1&0&0\\0&1&0\\0&0&0\\0&0&1\\1&1&0\\0&0&1
  \end{array} \right)
  &\rightarrow&
  (\proj{P}\x)' ~=~ (\proj{P'}{\x'})' ~=~ \left( \begin{array}{cc} 
    1&0\\0&0\\0&0\\0&1\\1&0\\0&1
  \end{array} \right)
\end{eqnarray*}
where $\x' = (010)$.

\subsubsection*{Affinification}

Let $\s$ be a string of $l$ bits, and from the matrix $P$ (\emph{resp.} the code $\cC$, the matroid $\cM$) we derive $P_\s$ (\emph{resp.} $\cC_\s$, $\cM_\s$) by deleting all rows that are orthogonal to $\s$, using the natural $\FF_2$ inner-product.  Note that this increases neither size (length) nor rank.  Note that $\cC_\s$ always contains the all-ones codeword, and (equivalently) $\cM_\s$ is always an \emph{affine} matroid, and indeed a \emph{minor} of $\cM$.

Here is a random example, with $n=6$, $r=3$, $l=4$, and $\s = (0110)$~:
\begin{eqnarray*}
  P ~=~ \left( \begin{array}{cccc} 
    1&1&0&1\\0&1&1&0\\0&0&0&0\\0&1&0&1\\1&0&1&1\\0&1&0&1
  \end{array} \right)
  &\rightarrow&
  P_\s ~=~ \left( \begin{array}{cccc} 
    1&1&0&1\\0&1&0&1\\1&0&1&1\\0&1&0&1
  \end{array} \right)
\end{eqnarray*}
In this example, the rank did not drop, but the length (size) of the code (matroid) dropped from 6 to 4.
Echelon-reduced forms of these example matrices are
\begin{eqnarray*}
  P' ~=~ \left( \begin{array}{ccc} 
    1&0&0\\0&1&0\\0&0&0\\0&0&1\\1&1&0\\0&0&1
  \end{array} \right)
  &\rightarrow&
  (P_\s)' ~=~ (P'_{\s'})' ~=~ \left( \begin{array}{ccc} 
    1&0&0\\0&1&0\\0&0&1\\0&1&0
  \end{array} \right)
\end{eqnarray*}
where $\s' = (101)$.  Note that $\x$ and $\s$ transform differently, to $\x'$ and $\s'$ respectively, under echelon-reduction.  This is because $\s$ really belongs to the \emph{dual} space, being the space of linear maps $\FF_2^l \rightarrow \FF_2$.  (It is convenient to think of $\x$ as a row-vector and $\s$ as a column-vector.)

\section{X-programs for $\IQP$}  \label{sect:main}
  
  The class $\mathbf{QNC}_f^0$ was introduced by H\o{}yer and Spalek \cite{lit:Hoy02} to capture the idea of allowing only a constant amount of scope for temporal complexity within a (uniform) family of circuits.  $\IQP$ goes `a step further' by allowing essentially \emph{no} temporal structure within the abstract quantum process.  The main simulation results for $\IQP$ distributions (as given by X-programs) and their marginals are given in \S\ref{sect:simulation} and \S\ref{sect:marginals}, respectively.

\subsection{Definition}

  The definition given below is a little different from that given in \cite{us:BJS}, to enable the combinatorial structure to be seen more clearly.

  An \emph{X-program $(P, \theta)$} gives a recipe for building a probability distribution.  Here $P$ denotes an $n$-by-$l$ matrix over $\FF_2$ and $\theta$ is some real angle.  The matrix $P$ is used to build a Hamiltonian of $n$ commuting terms on $l$ qubits, each term a product of Pauli X operators.  (Formally we demand that $l = O( poly(n) )$, though it turns out there is no point even taking $l \ge n$.)
\begin{eqnarray*}
  \H_P  &:=&  \sum_{a=1}^n \prod_{b=1}^l X_b^{P_{ab}}.
\end{eqnarray*}
Thus the columns of $P$ correspond to qubits, while the rows of $P$ correspond to `gates' (in a circuit) or `interactions' (in a Hamiltonian) on the qubits.

The distribution is then given by 
\begin{eqnarray*}
  \Pr[\X = \x]  &:=&  \Bigl| \bra\x \exp( i \theta \H_P ) \ket\0 \Bigr|^2,
\end{eqnarray*}
using the squares of the magnitudes of the transition amplitudes, as required by the Born rule for quantum processes.  We call this an $\IQP$ probability distribution.

\subsection{Implementation}

We consider that a quantum circuit for sampling from this distribution contains no `inherent temporal structure' because all terms in the Hamiltonian commute, and so it makes no difference (in principle) which is implemented first, or whether they are implemented simultaneously, somehow.  

In \cite{me:IQC,mythesis} we showed explicitly how to implement these circuits in certain theoretical architectures.  For example, one such implementation uses so-called \emph{graph states}, where $P$ is interpreted as (the biadjacency matrix of) a bipartite graph at whose vertices are located single qubits.  (A particularly interesting \emph{mild generalisation} of X-programs---still without introducing temporal structure---is to consider the preparation and \emph{arbitrary} measurement of graph states, with no intervening feed-forward/adaptive control of measurement results.)  But these implementation issues will not be of concern to us in the present analysis.  For this paper, the Hamiltonian $\H_P$ `takes priority' over any circuit that might be designed to implement it, or the unitary mappings it generates.

\subsection{Correlation coefficients}  \label{sect:correl}

  As well as being interested in $\IQP$ probability distributions and the individual probabilities $\Pr[\X = \x]$, we are also interested in the \emph{correlation coefficients}.  These are given by taking the $\FF_2$ Fourier transform of the \emph{probability vector}.  In symbols, we define
\begin{eqnarray*}
  \beta_\s  &:=&  2 \cdot \Pr[ \X \cdot \s = 0 ] - 1,
\end{eqnarray*}
and thence derive the following formul\ae{}~:
\begin{eqnarray*}
  \Pr[ \X \cdot \s = 0 ]  &=&  \frac{1+\beta_\s}2;  \\
  \Pr[ \X = \x ]          &=&  2^{-l} \cdot \sum_{\s \in \FF_2^l} (-1)^{\x \cdot \s} \cdot \beta_\s \\
                          &=&  \Ex_{\s}[~ (-1)^{\x \cdot \s} \cdot \beta_\s ~].
\end{eqnarray*}
The second line above shows that the probability distribution is entirely specified by the values $\beta_\s$, and therefore encodes $2^l-1$ degrees of freedom (the value $\beta_\0$ is fixed as 1 for all distributions).  Because a distribution encodes so much data, it can be a very cumbersome object to work with.

\subsection{Relation to matroids}

Here we give the theorems that show the link between probability distributions for X-programs and the weight enumerator.

\begin{theorem}  \label{thm:amplitude}
  For any binary matrix $P$ and any angle $\theta$,
\begin{eqnarray*}
  \bra\0 \exp( i\theta \H_P ) \ket\0  
    &=&  \alpha_{(P,\theta)}.
\end{eqnarray*}
Moreover, for $\x \not= \0$,
\begin{eqnarray*}
  \bra\x \exp( i\theta \H_P ) \ket\0  
    &=&  \alpha_{(\proj{P}\x,\theta)} - \alpha_{(P,\theta)},
\end{eqnarray*}
where $\proj{P}\x$ denotes the \emph{projection} defined in \S\ref{sect:trans}, and $\alpha$ is the function defined in \S\ref{sect:WETP}.
\end{theorem}

This theorem has not been published before.  We prove it in \S\ref{sect:proofs}.
\hfill  \qed

There is also a simple formula for the correlation coefficients $\beta_\s$, as follows.
\begin{theorem}  \label{thm:beta}
  Let $\s$ be a string of $l$ bits and let $(P,\theta)$ be an X-program on $l$ qubits.  Let $\beta_\s$ be the correlation coefficient associated to $\s$ for the associated $\IQP$ probability distribution.  This $\beta_\s$ is a real value, and can be specified in terms of the \emph{affinification} $P_\s$, as follows.
\begin{eqnarray*}
  \beta_\s  &=&  \alpha_{(P_\s,2\theta)}.
\end{eqnarray*}
\end{theorem}

We prove this theorem in \cite{me:IQC,mythesis}, and give a slightly simpler proof in \S\ref{sect:proofs}, for completeness.
\phantom{XXXXXXXXXX} \hfill  \qed

\subsection{Implications for simulation}  \label{sect:simulation}

In this section, we summarise a variety of observations about $\IQP$ probability distributions, for different values of $\theta$ and different classes of $P$.  (Later in \S\ref{sect:marginals}, we consider `breaking apart' $\IQP$ distributions into marginal probability distributions.)

\subsubsection*{$\theta = \frac\pi4$ (completely solved)}

  The case $\theta = \frac\pi4$ is interesting.  In that case, all values $\Pr[\X=\x]$ and $\Pr[\X\cdot\s=0]$ can be computed efficiently using Vertigan's algorithm, together with the theorems above.  Moreover, the actual probability distribution can be efficiently sampled classically, exactly.  This is because the unitary map $\exp( i \frac\pi4 \H_P )$ can be decomposed as a product of Clifford gates, whence the Gottesman-Knill theorem applies.  
  To see this, observe for example that 
\begin{eqnarray*}
  \exp\left( \frac{i\pi}4 \prod_{j=1}^{k} X_j \right) \cdot Z_1 \cdot \exp\left( -\frac{i\pi}4 \prod_{j=1}^k X_j \right)
    &=&  -i \cdot Z_1 \cdot \prod_{j=1}^k X_j,
\end{eqnarray*}
and so the Pauli group is fixed by this unitary.
  There is thus a link between the classical simulation of so-called \emph{stabiliser states}~\cite{lit:AG0406}, the fact that Clifford circuits are effectively devoid of temporal structure~\cite{lit:Browne06}, and Vertigan's algorithm for evaluating $T_\cM(-i,i)$ efficiently~\cite{lit:Vert98}.  
  
  The following is an analogue of the Gottesman-Knill Theorem for X-programs, giving a complete description of the probability distribution in this case~:
\begin{theorem}  \label{thm:sim_piby4}
For any X-program on $l$ qubits, with $\theta = \frac\pi4$, there is an efficiently computed \emph{affine space} $S$ of $\FF_2^l$ over which the associated probability distribution is \emph{supported} and \emph{uniform}.  That is to say, for some easily-determined $S$,
\begin{eqnarray*}
  \Pr[ \X = \x ]  &=&  2^{-dim(S)} \cdot \{ \x \in S \}.
\end{eqnarray*}
\end{theorem}

We prove this theorem in \S\ref{sect:proofs}.
\hfill  \qed

\subsubsection*{$\theta = \frac\pi8$ (main area of ongoing research)}

  Another interesting case---this time not so easy to simulate---occurs when $\theta = \frac\pi8$.  A quantum circuit could efficiently sample from the distribution, of course, but now a classical computer could efficiently determine any (polynomial number of) correlation coefficients of our choosing for that distribution, because $\beta_\s = \alpha_{(P_\s,\frac\pi4)}$ is efficiently computable by Vertigan's algorithm.  Therefore, a classical computer would be perfectly able to run an effective hypothesis test against a purported (sufficiently large) sample generated by the quantum circuit, \emph{regardless of the matrix $P$ chosen}.  Such an hypothesis test could be used to distinguish this `null' hypothesis from some appropriate `alternative' hypothesis.  This \emph{testing paradigm} generalises our earlier work \cite{me:IQC} and we anticipate it being useful in validating quantum computing implementations.
  We hope to explore the ramifications of this observation more fully in future work.
  
  The question then arises as to whether a classical computer could sample from the distribution by some other efficient means, or else sample efficiently from some statistically close distribution.  We know that $\BPP_{\mathbf{path}} \not= \PP$ (unless of course the Polynomial Hierarchy collapses), and so via Toda's theorem and the hardness of computing $\alpha_{(P,\frac\pi8)}$ \cite{lit:JVW90}, we can deduce that no classical algorithm will be able to sample efficiently from the $\IQP$ distribution exactly, for generic $P$.  In fact, we can use the theory of \emph{post-selection} to derive an independent proof (independent of \cite{lit:JVW90}) of the hardness of evaluating $W_\cC( e^{i\pi/4} )$, much as Aaronson \cite{lit:Aa04} used post-selection to derive an independent proof the closure properties of $\PP$ (see \cite{us:BJS} for a fuller account of this).  
  
  We leave open the possibility of an \emph{approximate} classical sampling technique.  It is important to determine just how close a classical algorithm can expect to get, with distance between distributions being measured as usual by the 1-norm (total variational distance)~: this is a subject we hope to address in future research.

\subsubsection*{Arbitrary $\theta$, but `special' $P$}

For \emph{arbitrary} $\theta$, Tutte polynomials can be evaluated for those matroids all of whose connected components are $O( \log n )$ in size, or indeed for \emph{graphic} matroids with $O( \log n )$ \emph{treewidth}~\cite{lit:And95}.  This suggests that somehow there is very limited `data flow' within an X-program computation.  

If $P$ is the \emph{incidence matrix} of a graph (\emph{i.e.} two 1s per row), then the formula of Proposition~\ref{propos:alpha} may be reinterpreted
\begin{eqnarray*}
  \alpha_{(P,\theta)} 
    &=&  \Ex_{\c}
         \left[~ \exp\bigl( i\theta \cdot ( n - 2|\c| ) \bigr) ~\right]
\end{eqnarray*}
so that $n$ counts the number of edges of this graph, and $\c$ ranges over all possible \emph{cuts}, with $|\c|$ denoting the weight of a cut (a \emph{cut} is simply a partition of vertices into two disjoint subsets, and its \emph{weight} is simply the number of edges that cross the partition).
Vertigan \cite{lit:Vert06} indicates that Kasteleyn's Theorem \cite{lit:Kas67} (which gives an efficient technique for counting perfect matchings on planar graphs) can be used to establish an explicit polytime algorithm for evaluating $\alpha_{(P,\theta)}$ whenever $P$ is the incidence matrix of a planar graph (\emph{cf.} \S\ref{sect:complexity}).  The algorithm for achieving this reduction is spelled out in more detail in \cite{lit:Fish66}.

\subsubsection*{Equivalence modulo $\theta$}

Before moving on, we give some insight as to why the values $\frac\pi4$ and $\frac\pi8$ are particularly noteworthy...

For two different X-programs $P$, $Q$, (with the same $\theta$ value), we say that $\H_P \sim_\theta \H_Q$ if they generate the same unitary map, $\exp( i\theta \H_P ) = \exp( i\theta \H_Q )$.  This notion of equivalence can be useful if one wishes to optimise certain properties of a matrix (perhaps for implementation reasons) without wishing to alter the distribution it would generate.
\begin{propos}
  If $\theta = c \cdot \pi/2^d$ for fixed integers $c$, $d$, then for any matrix $P$ (with $n$ rows and $l$ columns) it is computationally efficient to find a matrix $Q$ such that
\begin{itemize}
  \item
$\H_P  \sim_\theta  \H_Q$;
  \item
\# rows of $Q$ = $O(poly(n))$, \# columns of $Q$ = $l$;
  \item
each row of $Q$ has at most $d$ ones.
\end{itemize}
\end{propos}

To prove this, simply rewrite the Pauli operator $X_j$ as $1-2x_j$, where $x_j = \frac{1-X_j}2$ has integer eigenvalues.  Then rewrite $i\theta\H_P$ as a polynomial in the $x_j$s.  In this format, the number of terms may increase by an exponential factor (since a single term, \emph{e.g.} $X_1\ldots X_l$, may expand to form $2^l$ terms), but we can nonetheless efficiently list all those terms whose degree (number of $x_j$s) is at most $d$.  In the expansion, the remaining terms contribute nothing to $\exp(i\theta \H_P)$, because their coefficient is an integer multiple of $2\pi i$, and so we may ignore them.  For example
\begin{eqnarray*}
  \frac{i\pi}4 \cdot X_1 \cdot X_2 \cdot X_3
  &=& \frac{i\pi}4 \cdot (1-2x_1)(1-2x_2)(1-2x_3) \\
  &=& \frac{i\pi}4 \Bigl( 1 - 2x_1-2x_2-2x_3 + 4x_1x_2+4x_1x_3 + 4x_2x_3 \Bigr) - 2\pi i x_1x_2x_3 
\end{eqnarray*}
and so if $\theta=\frac\pi4$, we can replace $X_1X_2X_3$ by $X_1X_2X_3 + (1-X_1)(1-X_2)(1-X_3)$, which is of lower degree.

Therefore we take $i\theta\H_Q$ to be given by the polynomial $i\theta\H_P$ with high-degree terms (counting $x_j$s) removed.  Rewriting this as a polynomial in $X_j$s, we can recover $Q$ with the desired properties.
\hfill  \qed 

As a consequence of this, for studying $\theta=\frac\pi4$ we can make do with limiting to the class of \emph{graphic} matroids, but for $\theta=\frac\pi8$ we should also allow matrices with three 1s per row.  It is doubtful that one can infer that graphic matroids are \emph{necessarily} trivial in this context, since hardness results are shown in \cite{us:BJS} where the underlying matroids are all graphic (and $\theta = \frac\pi8$).  Nonetheless, the number of \emph{genuinely different} programs possible for a given $n$ would seem to be smaller for $\theta = \frac\pi4$ or even for $\theta = \frac\pi8$ than for, say, $\frac\pi3$, because of this effective restriction on row density.

\subsection{Simulating marginal distributions}  \label{sect:marginals}

What happens if we attempt to simulate only a few of the output bits from an X-program, assuming that the remaining bits from $\X$ are \emph{traced out}?  For this, we need some additional concepts.
Let $m$ be a linear idempotent map on $\FF_2^l$, so $m=m^2$.  This is called a \emph{projector}, and we do not assume it to be orthogonal.  For a (primal) vector $\x$, if $m(\x)=\x$ then we say that $\x$ is \emph{supported by} $m$; equivalently we say that $\x$ is \emph{in the range} of $m$.  Let $|m|$ denote the dimension of the range of $m$, so $l-|m|$ denotes the dimension of its kernel. 
Let $K$ and $R$ denote the kernel and range of $m$, and let $K^*$ and $R^*$ denote the kernel and range of its dual ($m^*$), so that $K^*$ is the perpendicular space to $R$, and $R^*$ is the perpendicular space to $K$.  (This is because if $\x=m(\x)$ then $\y\cdot\x = \y\cdot m(\x) = m^*(\y) \cdot \x$, and so if also $m^*(\y)=\0$, then $\y\cdot\x=0$, \emph{i.e.} $\x \bot \y$, and so on.)  

We say that \emph{$m$ is supported on $b$ (qu)bits} if $K$ contains $l-b$ distinct vectors of Hamming weight 1.  Note that it is possible for the number of bits $b$ on which $m$ is supported to be larger than the rank $|m|$.  
Likewise, $\x$ might be supported by $m$ even if $|\x| > |m|$.
But if the matrix representation of $m$ were diagonal, then $b=|m|$ and we could think of $m$ simply as \emph{masking off} certain bit-locations that are to be traced out. 

Define a \emph{marginal} probability distribution via the random variable $m(\X)$, according to the natural construction
\begin{eqnarray*}
  \Pr[ m(\X) = \x ] 
    &:=& \{ \x \in R \} \cdot \sum_{ \k \in K } \Pr[ \X = \x+\k ].
\end{eqnarray*}

\begin{propos}  \label{propos:marginal}
  Let $\X$ be a random variable for the distribution of an X-program $(P,\theta)$ and let $m$ be a projector whose range is denoted $R$ and for which the range of $m^*$ is denoted $R^*$.  Restricted to $R$, the random variable $m(\X)$ has probability distribution
\begin{eqnarray*}
  \Pr[ m(\X) = \x ] 
    &=& \Ex_{\s \in R^*}[~ (-1)^{\x\cdot\s} \cdot \alpha_{(P_\s,2\theta)} ~].
\end{eqnarray*}
\end{propos}

This is seen using the following chain of reasoning, working from the definition above, recalling \S\ref{sect:correl} and Theorem~\ref{thm:beta}.
\begin{eqnarray*}
  \Pr[ m(\X) = \x ] 
    &=& \{ \x \in R \} \cdot \sum_{ \k \in K } 2^{-l} \cdot \sum_{\s \in \FF_2^l} (-1)^{\x\cdot\s + \k\cdot\s} \cdot \beta_\s \\
    &=& 2^{-l} \cdot \{ \x \in R \} \cdot \sum_{\s \in \FF_2^l} (-1)^{\x\cdot\s} \cdot \beta_\s \cdot \sum_{ \k \in K }(-1)^{\k\cdot\s} \\
    &=& 2^{-|m|} \cdot \{ \x \in R \} \cdot \sum_{\s \in R^*} (-1)^{\x\cdot\s} \cdot \beta_\s \\
    &=& \{ \x \in R \} \cdot \Ex_{\s \in R^*}[~ (-1)^{\x\cdot\s} \cdot \alpha_{(P_\s,2\theta)} ~].
\end{eqnarray*}
\hfill  \qed

Such marginal distributions are often used in defining \emph{decision languages} from families of distributions arising from uniform families of computational circuits \cite{us:BJS}.  
Marginal distributions are also useful when attempting to render classical sampling algorithms, because it is clear that if each probability of each marginal distribution is computable, then there are many ways in which samples from the distribution can be simulated efficiently.

\subsubsection*{Other combinatorial results (strong simulation of marginals)}

A theorem about $\theta = \frac\pi8$ that works for arbitrary $P$~:-

\begin{theorem}  \label{thm:piby8}
  For an X-program $(P,\theta)$ with $n$ terms and angle $\theta = \frac\pi8$, giving rise to output distribution $\X$, for any `masking' projector $m$ with $|m| = O(\log(n))$, the probability vector for the marginal distribution $m(\X)$ is polynomially long and can be explicitly computed (classically) efficiently, regardless of $P$.
\end{theorem}

A theorem for `sparse' $P$ that works for arbitrary $\theta$~:-

\begin{theorem}  \label{thm:sparse}
  For an X-program $(P,\theta)$ with $n$ terms and arbitrary angle $\theta$, giving rise to output distribution $\X$, for any `masking' projector $m$ with $|m| = O(\log(n))$ \emph{that is actually supported on $O(\log n)$ of the qubits}, the probability vector for the marginal distribution $m(\X)$ is polynomially long and can be explicitly computed (classically) efficiently, provided there is some constant $c$ (independent of the problem instance) bounding the number of 1s in any column of $P$.
\end{theorem}

A theorem for `graphic' $P$ that works for arbitrary $\theta$~:-

\begin{theorem}  \label{thm:graphic}
  For an X-program $(P,\theta)$ with $n$ terms and arbitrary angle $\theta$, giving rise to output distribution $\X$, for any `masking' projector $m$ with $|m| = 2$ \emph{that is actually supported on at most $2$ of the qubits}, the probability vector for the marginal distribution $m(\X)$ can be explicitly computed (classically) efficiently, provided each row of $P$ contains at most two 1s. 
\end{theorem}

Elementary proofs are given in \S\ref{sect:proofs}.
\hfill  \qed

\subsubsection*{Non-combinatorial results (weak simulation of marginals)}

Moreover, \emph{without considering Tutte polynomials at all}, when $|m|=1$, and say $R = \{ \0, \m \}$, $R^* = \{ \0, \m^* \}$, we can always sample classically from the distribution $m(\X)$ over 1-bit strings---\emph{i.e.} emulate single output qubits independently---because of the formula
\begin{eqnarray*}
  \Pr[ m(\X) = \0 ] 
    &=&  \Pr[ \X \cdot \m^* = 0 ]  \\
    &=&  \frac{1+\alpha_{(P_{\m^*},2\theta)}}2  \\
    &=&  \frac{1+\Ex_{\c \in \cC(P_{\m^*})}
         \left[~ \exp\bigl( 2i\theta \cdot ( n_{\m^*} - 2|\c| ) \bigr) ~\right]}2 \\
    &=&  \Ex_{\c\in\cC(P_{\m^*})}[~ \cos^2(~ \theta \cdot (n_{\m^*} - 2|\c|) ~) ~],  \\
  \Pr[ m(\X) = \m ] 
    &=&  \Ex_{\c\in\cC(P_{\m^*})}[~ \sin^2(~ \theta \cdot (n_{\m^*} - 2|\c|) ~) ~],
\end{eqnarray*} 
where $n_{\m^*}$ denotes the length of $\cC( P_{\m^*} )$ as usual.
Note the use of Proposition~\ref{propos:alpha} in this deduction, and the fact that $\alpha_{(P_{\m^*},2\theta)}$ is real.
We can render this sample classically efficiently by sampling $\cC(P_{\m^*})$ uniformly, even though we need have no idea how to \emph{compute} the value~$\alpha_{(P_{\m^*},2\theta)}$.

This result can be generalised to the following theorem~:
\begin{theorem}  \label{thm:sample}
  For an X-program ($n$ terms by $l$ qubits) with any $\theta$, if $m$ is a projector on $\FF_2^l$ inducing a marginal probability distribution, with $|m| = O(\log(n))$ there is then an efficient (i.e. $O(poly(n))$ time) classical algorithm for sampling from this marginal distribution, accurate to exponential precision.
\end{theorem}

This is proved in \cite{us:BJS} without direct reference to combinatorics.  We also provide a proof below in \S\ref{sect:proofs} without direct reference to quantum computing.
\hfill  \qed

\section{Appendix of Proofs}  \label{sect:proofs}

The proofs of Theorems \ref{thm:amplitude}, \ref{thm:beta}, \ref{thm:sim_piby4}, \ref{thm:piby8}, \ref{thm:sparse}, \ref{thm:graphic}, \ref{thm:sample} are given in this section.

\subsubsection*{Proof of Theorem~\ref{thm:amplitude}}

This Theorem directly links probabilities for X-programs to Tutte polynomials.  

  First we make a change of basis, writing
\begin{eqnarray*}
   \bra\0 \exp( i \theta \H_P ) \ket\0   
     &=&  \bra+ \exp\left( i \theta \sum_{a=1}^n \prod_{b=1}^l Z_b^{P_{ab}} \right) \ket+
.
\end{eqnarray*}
Because this contains a matrix that is patently diagonal, and because its vectors $\bra+$ and $\ket+$ represent uniform superpositions (with matching phase), we can assert that this expression evaluates to
\begin{eqnarray*}
  2^{-l} \cdot Tr\left[~ \exp\left( i \theta \sum_{a=1}^n \prod_{b=1}^l Z_b^{P_{ab}} \right) ~\right]
  &=&  2^{-l} \cdot \sum_{\s \in \FF_2^l} \exp\left( i \theta \sum_{a=1}^n \prod_{b=1}^l (-1)^{P_{ab} \cdot s_b} \right) \\
  &=&  \Ex_{\s \in \FF_2^l}\left[~ \exp\left( i \theta \sum_{a=1}^n (-1)^{(P \cdot \s)_a} \right) ~\right] \\
  &=&  \Ex_{\c \in \cC(P)} \left[~ \exp\left( i \theta \sum_{a=1}^n (-1)^{c_a} \right) ~\right]\\
  &=&  2^{-r} \cdot \sum_{\c \in \cC(P)} \exp\Bigl( i \theta ( n - 2|\c| ) \Bigr) \\
  &=& \alpha_{(P,\theta)}.
\end{eqnarray*}

To evaluate the other transition amplitudes, we need to observe that 
\begin{eqnarray*}
  \bra\x  &=&  \bra\0 \exp\left( i \frac\pi2 \left( -1 + \prod_{b=1}^l X_b^{x_b} \right) \right).
\end{eqnarray*}
The case $\theta = \frac\pi{2t}$ is most easily dealt with, where $t \in \ZZ^+$.
Then let $Q(j)$ denote a matrix obtained from $P$ by appending $j$ extra rows identical to $\x$, and we quickly see that
\begin{eqnarray*}
   \bra\x \exp\left( i \theta \sum_{a=1}^n \prod_{b=1}^l X_b^{P_{ab}} \right) \ket\0   
     &=&  e^{-i \theta t} \cdot \bra\0 \exp\left( i \theta \sum_{a=1}^{n+t} \prod_{b=1}^l X_b^{Q(t)_{ab}} \right) \ket\0 \\
     &=&  -i \cdot \alpha_{(Q(t),\theta)}.
\end{eqnarray*}
Now using the fact that $\alpha_{(Q(j),\theta)}$ is proportional to the evaluation of some Tutte polynomial, we can use the deletion and contraction formul\ae{} to remove the extra rows that were added, according to the following chain of reasoning.  Let $R(j)$ denote the matrix $\proj{P}\x$ with $j$ extra loops (all-zero rows) appended.  This corresponds to the matroid obtained by contracting one of the appended rows from $Q(j+1)$.  

There are two sub-cases to consider.  In the first, the size and rank of $\cM(Q(t))$ are $n+t$ and $r$ respectively, and $\x$ lies within the span of $E$, and $\cC(\proj{P}\x)$ is strictly contained within $\cC(P)$.
We deduce
\begin{eqnarray*}
  T_{\cM(Q(j))}( x, y )  &=&  T_{\cM(Q(j-1))}( x, y ) ~+~ T_{\cM(R(j-1))}( x, y ), \\
  T_{\cM(R(j))}( x, y )  &=&  y^j \cdot T_{\cM(\proj{P}\x)}( x, y ),
\end{eqnarray*}
and hence
\begin{eqnarray*}
  T_{\cM(Q(t))}( x, y )  &=&  \frac{1-y^t}{1-y} \cdot T_{\cM(\proj{P}\x)}( x, y ) ~+~ T_{\cM(P)}( x, y ).
\end{eqnarray*}
Substituting this into our formula for $\alpha$, and setting $x = \frac{e^{i\theta}+e^{-i\theta}}{e^{i\theta}-e^{-i\theta}}$ and $y = e^{2i\theta}$, we obtain
\begin{eqnarray*}
  -i \cdot \alpha_{(Q(t),\theta)}
    &=&  -i \cdot e^{i \theta (r-n-t)} \cdot i^{r} \cdot \sin^{r}\theta \cdot \left( \frac{1-e^{2i\theta t}}{1-e^{2i\theta}} \cdot T_{\cM(\proj{P}\x)}( x, y ) ~+~ T_{\cM(P)}( x, y ) \right) \\
    &=&  -e^{i \theta (r-n)} \cdot i^{r} \cdot \sin^{r}\theta \cdot \left( \frac{2}{1-e^{2i\theta}} \cdot T_{\cM(\proj{P}\x)}( x, y ) ~+~ T_{\cM(P)}( x, y ) \right) \\
    &=&  e^{i \theta (r-1-n)} \cdot i^{r-1} \cdot \sin^{r-1}\theta \cdot T_{\cM(\proj{P}\x)}( x, y ) \\
    &&~~~-~ e^{i \theta (r-n)} \cdot i^{r} \cdot \sin^{r}\theta \cdot  T_{\cM(P)}( x, y ) \\
    &=&  \alpha_{(\proj{P}\x,\theta)} ~-~ \alpha_{(P,\theta)}. 
\end{eqnarray*}

The second sub-case has that the size and rank of $\cM(Q(t))$ are $n+t$ and $r+1$ respectively, and $\x$ does not lie within the span of $E$, and $\cC(\proj{P}\x) = \cC(P)$.
We deduce
\begin{eqnarray*}
  T_{\cM(Q(1))}( x, y )  &=&  x \cdot T_{\cM(P)}( x, y ), \\
  T_{\cM(Q(j))}( x, y )  &=&  T_{\cM(Q(j-1))}( x, y ) ~+~ T_{\cM(R(j-1))}( x, y ), \\
  T_{\cM(R(j))}( x, y )  &=&  y^j \cdot T_{\cM(\proj{P}\x)}( x, y ),
\end{eqnarray*}
and hence
\begin{eqnarray*}
  T_{\cM(Q(t))}( x, y )  &=&  \left( \frac{1-y^t}{1-y} + x - 1 \right) \cdot T_{\cM(P)}( x, y ).
\end{eqnarray*}
On substituting in $x = \frac{e^{i\theta}+e^{-i\theta}}{e^{i\theta}-e^{-i\theta}}$ and $y = e^{2i\theta}$, we see that the expression vanishes.  
It therefore still holds (now somewhat vacuously) that
\begin{eqnarray*}
  -i \cdot \alpha_{(Q(t),\theta)}
    &=&  \alpha_{(\proj{P}\x,\theta)} ~-~ \alpha_{(P,\theta)}. 
\end{eqnarray*}

With both sub-cases established, we argue that since this equation holds whenever $\theta = \frac\pi{2t}$ for $t \in \ZZ^+$, so it must hold for all $\theta$.  This is true because the expressions can be thought of as Laurent polynomials in $z=e^{2i\theta}-1$ (\emph{i.e.} elements of $\CC[z,z^{-1}]$, because $y=z+1$ and $x=1+2z^{-1}$), and if two Laurent polynomials agree at an infinite number of distinct places, then they must agree everywhere. 
\hfill  \qed

\subsubsection*{Proof of Theorem~\ref{thm:beta}}

The proof of this Theorem first appears in \cite{me:IQC}, though the present version is a little simpler.

We start with the basic definition of $\Pr[\X=\x]$, and change into the diagonal basis in order to make the summation and lose the \emph{bra-ket} notation.  Then we work on removing the modulus signs.  This gives 
\begin{eqnarray*} 
  \Pr(\X=\x)
  &=& \left| \bra\x \exp\left(~i \theta \sum_{a=1}^n \prod_{b=1}^l X_b^{P_{ab}}~\right) \ket{\0} \right|^2 \\
  &=& \left| 2^{-l}\sum_{\a \in \FF_2^l} (-1)^{\x \cdot \a}\bra\a \exp\left(~i \theta \sum_{a=1}^n \prod_{b=1}^l Z_b^{P_{ab}}~\right) \sum_{\b \in \FF_2^l} \ket{\b} \right|^2 \\
  &=& \left| 2^{-l}\sum_{\a \in \FF_2^l} (-1)^{\x \cdot \a} \exp\left(~i \theta \sum_{a=1}^n (-1)^{P_{a} \cdot \a}~\right) \right|^2 \\
  &=& \left| ~\Ex_{\a} \left[ (-1)^{\x \cdot \a} \exp\left(~i\theta \sum_{a=1}^n (-1)^{P_a \cdot \a} ~\right) \right] ~\right|^2 \\
  &=& \Ex_{\a,\d} \left[ (-1)^{\x \cdot \d} \exp\left(~ i\theta \sum_{a=1}^n (-1)^{P_a \cdot \a} \Bigl(1 - (-1)^{P_a \cdot \d}\Bigr) ~\right) \right].  
\end{eqnarray*}

Next, we take the definition for $\beta_\s$ in terms of the Fourier transform, drop in the formula above, and then just kill off the spare variables.
\begin{eqnarray*} 
  \beta_\s
    &=& \sum_{\x \in \FF_2^l} (-1)^{\x \cdot \s} \cdot \Pr( \X = \x )  \\
    &=& \sum_{\x \in \FF_2^l} (-1)^{\x \cdot \s} \cdot \Ex_{\a,\d} \left[ (-1)^{\x \cdot \d} \exp\left(~ i\theta \sum_{a=1}^n (-1)^{P_a \cdot \a} \Bigl(1 - (-1)^{P_a \cdot \d}\Bigr) ~\right) \right] \\
    &=& \Ex_{\a,\d} \left[ 2^l \cdot \{\s=\d\} \cdot \exp\left(~ i\theta \sum_{a=1}^n (-1)^{P_a \cdot \a} \Bigl(1 - (-1)^{P_a \cdot \d}\Bigr) ~\right) \right] \\
    &=& \Ex_{\a} \left[ \exp\left(~ i\theta \sum_{a=1}^n (-1)^{P_a \cdot \a} \Bigl(1 - (-1)^{P_a \cdot \s}\Bigr) ~\right) \right].
\end{eqnarray*}

Now it is clear that whenever $P_a \cdot \s = 0$, then the corresponding 
term vanishes, independent of the variable $\a$.  Thus we need only include in the sum those terms for which $P_a \cdot \s = 1$.
\begin{eqnarray*} 
  \beta_\s
    &=& \Ex_{\a \in \FF_2^l} \left[ \exp\left(~ 2i\theta \sum_{P_a \cdot \s = 1} (-1)^{P_a \cdot \a} ~\right) \right].
\end{eqnarray*}

This formula we can naturally express in terms of the code $\cC(P_\s)$.  Let $n_\s$ be the size of the affinified matroid (equivalently, the length of the resulting code).  Use Proposition~\ref{propos:alpha} to finish the deduction.
\begin{eqnarray*} 
  \beta_\s
    &=& \Ex_{\c \in \cC(P_\s)} \left[~ \exp\left(~ 2i\theta (n_\s - 2|\c|) ~\right) ~\right] \\
    &=& \alpha_{(P_\s, 2\theta)}.
\end{eqnarray*}
\vskip -0.8cm
\hfill  \qed

\subsubsection*{Proof of Theorem \ref{thm:sim_piby4}}

This Theorem and its proof are essentially a reworking of Vertigan's algorithm for $T_\cM( -i, i )$ on binary matroids (because of Theorem~\ref{thm:amplitude}), but here expressed more directly in terms of X-programs and their distributions.  (Note that Vertigan's algorithm \cite{lit:Vert98} effectively computes $\alpha_{(P,\frac\pi4)}$ for input $P$, whereas the algorithm presented below discovers probabilities, and hence only $|\alpha_{(P,\frac\pi4)}|$.)

Let $\x$ be any element of $\FF_2^l$ and consider
\begin{eqnarray*}
  \Pr[ \X = \x ]       
    &=&  \Ex_{\s \in \FF_2^l}[~ (-1)^{\x \cdot \s} \cdot \alpha_{(P_\s,\frac\pi2)} ~] \\
    &=&  \Ex_{\s \in \FF_2^l}[~ (-1)^{\x \cdot \s} \cdot 2^{-r_\s} \cdot i^{n_\s} \cdot W_{\cC(P_\s)}( -1 ) ~].
\end{eqnarray*}

The expression $2^{-r_\s} \cdot W_{\cC(P_\s)}( -1 )$ has some particularly simple interpretations.  Evidently, if $\cC( P_\s )$ is an even code, then this expression takes the value 1; otherwise it vanishes.  But the following conditions are all equivalent~:
\begin{itemize}
  \item
$\cC( P_\s )$ is an even code;
  \item
the column vector $P \cdot \s$ is orthogonal to every codeword of $\cC(P)$;
  \item
$(P \cdot \s)^T \cdot P$ is all-zero;
  \item
$\s \in Ker( P^T \cdot P )$.
\end{itemize}  
Accordingly, we can restrict attention to those $\s \in Ker( P^T \cdot P )$.  This is simply a linear constraint on $\s$, because the kernel of a linear map is a subspace of its domain.  Write $V$ as short-hand for $Ker(P^T \cdot P)$.  Observing that $n_\s$ is given by $|P\cdot \s|$ (the Hamming weight of the column vector $P \cdot \s$), we see that it is indeed always even in the case $\s \in V$, as one might expect.

Then define the subspace $U := \{~ \s ~:~ \s \in V,~ i^{|P\cdot \s|} = 1 ~\} \le V$, consisting of those elements $\s$ of $V$ for which 4 divides $n_\s$.  (Note that $U$ really is a subspace, because it is closed under addition of elements, because for all $\s \in V$, $P \cdot \s$ is orthogonal to everything of the form $P \cdot \a$.)

Combining these ideas,
\begin{eqnarray*}
  \Pr[ \X = \x ]       
    &=&  2^{-l} \cdot \sum_{\s \in \FF_2^l} (-1)^{\x \cdot \s} \cdot 2^{-r_\s} \cdot i^{n_\s} \cdot W_{\cC(P_\s)}( -1 ) \\
    &=&  2^{-l} \cdot \sum_{\s \in V} (-1)^{\x \cdot \s} \cdot i^{n_\s} \\
    &=&  2^{-l} \cdot \left( 2\sum_{\s \in U} (-1)^{\x \cdot \s} ~-~ \sum_{\s \in V} (-1)^{\x \cdot \s} \right).
\end{eqnarray*}

Now this can be simplified further, because each of these latter sums either vanishes or else is equal to the number of terms it contains.  
Finally, we consider this expression in two different cases.  

First case~:- If $U = V$ (because 4 divides $n_\s$ for every $\s \in V$) then the expression simplifies to 
\begin{eqnarray*}
  \Pr[ \X = \x ]       
    &=&  2^{-l} \cdot \left( \sum_{\s \in V} (-1)^{\x \cdot \s} \right) \\
    &=&  2^{dim(V)-l} \cdot \{~ \x ~\bot~ V ~\},
\end{eqnarray*}
which means that the distribution is uniform over the subspace of vectors orthogonal to the kernel of $P^T\cdot P$.  We could naturally denote this `support' set $S_1 := V^\bot$.  Because $\0 \in V^\bot$, this case leads to $|\alpha_{(P,\frac\pi4)}| = 2^{\frac{dim(V)-l}2}$.

Second case~:- If $U < V$ (because only half of the elements of $V$ have that 4 divides $n_\s$) then the expression simplifies to
\begin{eqnarray*}
  \Pr[ \X = \x ]       
    &=&  2^{dim(V)-l} \cdot \bigl( \{~ \x ~\bot~ U ~\} ~-~ \{~ \x ~\bot~ V ~\} \bigr)  \\
    &=&  2^{dim(V)-l} \cdot \{~ \x ~\bot~ U ~\} \cdot \{~ \x ~\not\!\!\bot~ V ~\},
\end{eqnarray*}
which means that the distribution is uniform over the affine space comprising the (unique) non-trivial coset of $V^\bot$ in $U^\bot$, equivalently denoted $S_2 := U^\bot \backslash V^\bot$.  Because $\0 \not\in U^\bot \backslash V^\bot$, this case leads to $\alpha_{(P,\frac\pi4)} = 0$.  Considering our example $P$ from \S\ref{sect:WETP}~: $l=4$, $V = \left< (1001), (0111) \right>$, $U = \left< (0111) \right>$, and indeed $W_{\cC(P)}(-i) = 0$.

The two cases ($S_1$, $S_2$) are distinguished by computing (bases for) $V$ and $U$.  It is clear that $V$ is readily computed.  To find $U$ (and thus establish the proof) one need only examine a basis for $V$.  For a fixed basis, if every basis element $\s$ of $V$ has that 4 divides $n_\s$, then by closure it must be that $U = V$.  Alternatively, if some `bad' subset of the basis has elements $\s$ for which 4 does not divide $n_\s$, then by a simple elimination technique we can change the basis (adding one `bad' element into another to make it `good') so that exactly one basis element has 4 not dividing $n_\s$.  With this one `bad' element removed, the remaining set forms a basis for $U$.
\hfill  \qed

\subsubsection*{Proof of Theorem \ref{thm:piby8}}

Certainly when $|m| = O( \log n )$, the number of degrees of freedom associated to the distribution for $m(\X)$ scales only polynomially in $n$.  

If also $\theta = \frac\pi8$ then the entirety of this distribution becomes classically computable and simulable.  This is achieved by running Vertigan's algorithm a polynomial number of times to compute the $\beta_\s$ values explicitly for all $\s \in R^*$, in accordance with Proposition~\ref{propos:marginal}~:
then we compute the Fourier transform of that vector to find the probability vector for the distribution in question.
\hfill  \qed

\subsubsection*{Proof of Theorems~\ref{thm:sparse} and \ref{thm:graphic}}

As above, $|m| = O( \log n )$ implies that we shall need to compute only polynomially many correlation coefficients ($\beta_\s$).  Moreover, the condition on $m$ actually being supported on $O(\log n)$ of the qubits (for Theorem~\ref{thm:sparse}) means that we need compute $\beta_\s$ only for cases where $|\s| = O(\log n)$.  Recall (Theorem~\ref{thm:beta} \& Proposition~\ref{propos:Greene}) that $\beta_\s$ depends on the Tutte polynomial, rank, and size of $\cM( P_\s )$.

For any $\s \in \FF_2^l$ with $|\s| = O(\log n)$, the size (and hence rank) of $\cM( P_\s )$ evidently cannot be bigger than $c \cdot O( \log n )$, where $c$ bounds the number of 1s in every column of $P$ (this can be seen \emph{e.g.} by bounding the number of 1s there can be within the columns indicated by $\s$, and each row of $P_\s$ must contain at least one of these 1s).  
Even a `dumb' (exponentially slow) algorithm for evaluating Tutte polynomials will require only time scaling polynomially in $n^c$ for such tiny matroids.  Hence all the required correlation coefficients, and thence the entire marginal probability distribution, can be computed as required for Theorem~\ref{thm:sparse}.

Note that neither the bounds on the Hamming weight of rows/columns of $P$ nor the conditions regarding $m$ being supported on some number of qubits are matroid invariants~: rather, they depend critically on the actual presentation of $P$ and $m$ and are not preserved under basis transformation.

For Theorem~\ref{thm:graphic}, we have at most three correlation coefficients to compute (since the range of $m$ has dimension at most two, and hence cardinality at most 4, but $\beta_\0$ is always 1).  In each case, $\cM( P_\s )$ is the cycle matroid of some bipartite graph one of whose partitions contains at most two vertices (corresponding to the qubits selected by $m$).  It is relatively straightforward to come up with a polytime algorithm for evaluating the Tutte polynomial of such an easy graph.  

We will not bore the reader with an explicit derivation of such an algorithm. 
Suffice it to say that it is relatively straightforward to identify and group the graph minors intelligently, so as to make recursive evaluation of the Tutte polynomial efficient.  This is because every minor of such a graph---having once had all its coloops deleted---is never more than two edge-contractions away from a `star-shaped' graph, whose Tutte polynomial takes the explicit form
\begin{eqnarray*}
  T_{\mbox{star}(\a)}(x,y) &=& \prod_{j=1}^k \left( \frac{y^{a_j}-1}{y-1} + x - 1 \right),
\end{eqnarray*}
for some integer tuple $\a$ describing the star-graph.
It is likely that this sort of construction could be generalised (presumably in terms of \emph{treewidth}), extending the Theorem to cope with $|m|>2$ (at the cost of much ink), but no such analysis has yet been made.
\phantom{XXXXXXXXXXXXXXXX}\hfill  \qed

\subsubsection*{Proof of Theorem \ref{thm:sample}}

  We already saw (\emph{cf.} Proposition~\ref{propos:marginal}, Theorem~\ref{thm:beta}, and Proposition~\ref{propos:alpha}) that for any projector $m$ on $\FF_2^l$ (range denoted $R$), for $\x$ restricted to $R$,
\begin{eqnarray*}
  \Pr[m(\X)=\x]
  &=& \Ex_{\s \in R^*}\left[~ (-1)^{\x\cdot\s} \cdot \beta_\s ~\right] \\
  &=& \Ex_{\s \in R^*}\left[~ (-1)^{\x\cdot\s} \cdot \Ex_{\c \in \cC(P_\s)}\left[~ \cos\bigl( 2\theta \cdot (n_\s - 2|\c|) \bigr) ~\right] ~\right],
\end{eqnarray*} 
where $n_\s$ denotes the length of $\cC(P_\s)$ as usual.  
Conceptually breaking the code $\cC(P_\s)$ into the direct sum of two pieces across the boundary created by the mask $m$, we write the codeword $\c \in \cC(P_\s)$ as the sum of $P_\s \cdot \t$ and $P_\s \cdot \k$, for $\t$ and $\k$ in $R^*$ and $K^*$, respectively.  Hence we find that another way to write the same distribution is
\begin{eqnarray*}
  \Pr[m(\X)=\x]
  &=& \Ex_{\k \in K^*} \Ex_{\s,\t \in R^*} \left[~ (-1)^{\x\cdot\s} \cdot \cos\Bigl( 2\theta \cdot \bigl(n_\s - 2|P_\s \cdot (\t+\k)|\bigr) \Bigr) ~\right].
\end{eqnarray*}

Therefore, for any $\k \in K^*$, let us define
\begin{eqnarray*}
  \Pr[~ m(\X)=\x ~|~ \k ~]
  &:=& \Ex_{\s,\t \in R^*} \left[~ (-1)^{\x\cdot\s} \cdot\cos\Bigl( 2\theta \cdot \bigl(n_\s - 2|P_\s \cdot (\t+\k)|\bigr) \Bigr) ~\right].
\end{eqnarray*} 

Observe that this expression may be rewritten as follows.
\begin{eqnarray*}
  \Pr[~ m(\X)=\x ~|~ \k ~]
  &=&  \left|~ \Ex_{\t \in R^*}\left[~ (-1)^{\x\cdot\t} \cdot \exp\left( i\theta \cdot \sum_{j=1}^n (-1)^{P_j \cdot (\t+\k)} \right) ~\right] ~\right|^2.
\end{eqnarray*} 
(The computation proceeds along similar lines to the one in the proof of Theorem~\ref{thm:beta}, but in reverse.)
And from here we see that this genuinely describes a probability distribution, \emph{i.e.} is non-negative for all $\x$.

It then follows immediately that the distribution for $m(\X)$ is just the uniform combination of the distributions above, over $\k \in K^*$.  
Moreover, when $|m| = O(\log(n))$, the dimension of the range $R^*$ is sufficiently small that there are only polynomially many $\t$ to worry about.
To sample from $m(\X)$ one need therefore only choose $\k$ uniformly from $K^*$ and then explicitly compute the entire probability vector shown above for that value of $\k$, to an appropriate level of accuracy.  Provided we use a fresh vector $\k$ for each sample, the simulation will be exact, up to accuracy of the numerical precision of the real computation.
\hfill  \qed

\end{document}